# BOTH GENERIC DESIGN AND DIFFERENT FORMS OF DESIGNING

WILLEMIEN VISSER

*INRIA - National Institute for Research in Computer Science and Control*
*EIFFEL - Cognition & Cooperation in Design*
*Rocquencourt*
*78153 LE CHESNAY CEDEX (FRANCE)*

*tel: + 33 (0) 1 39 63 52 09*
*fax: + 33 (0) 1 39 63 59 95*
*email: Willemien.Visser@inria.fr*

**Abstract.** This paper defends an augmented cognitively oriented "generic-design hypothesis": There are both significant similarities between the design activities implemented in different situations and crucial differences between these and other cognitive activities; yet, characteristics of a design situation (i.e., related to the designers, the artefact, and other task variables influencing these two) introduce specificities in the corresponding design activities and cognitive structures that are used. We thus combine the generic-design hypothesis with that of different "forms" of designing. In this paper, outlining a number of directions that need further elaboration, we propose a series of candidate dimensions underlying such forms of design.

## 1. Introduction

"Activities as diverse as software design, architectural design, naming and letter-writing appear to have much in common" is, based on studies of these diverse activities, one of Thomas and Carroll (1979/1984)'s conclusions. In his paper "Comparison of design and nondesign problem spaces," Goel (1994) analyses the similarities between the cognitive processes involved in different design problem-solving tasks (architecture, mechanical engineering, and instructional design) and the differences between these design tasks and nondesign problem-solving tasks (cryptarithmetic and the Moore-Anderson logic task). Together with an older paper co-authored by Pirolli (Goel & Pirolli, 1989), this Goel (1994) publication is often referred to by researchers from various domains of design in order to present the "generic-design" principle, which, according to Goel (1994), "was one of the corner stone premises of the design methodology movement" (p. 53). Whereas this movement was not based on an analysis of the cognitive activities underlying design, Goel and Pirolli (1989) aimed to "motivate the notion of generic design within information-processing theory," that is, within the framework that Newell and Simon (1972) developed for the cognitive analysis of problem-solving activities. Goel and Pirolli indeed differentiated design problem solving from nondesign problem solving "by identifying major invariants in the design problem space"—on the basis, however, of tasks that were rather artificial (especially the nondesign tasks) and performed in artificially restricted laboratory conditions (all tasks).

Using Google with the keyword "generic design" mainly leads to references in the domains of AI and knowledge-acquisition (e.g., KADS and successor work), based, for example, on Chandrasekaran (1983)'s "generic tasks," on the notion of "generic design methods," and other "generic design agents." These are thus all normatively based approaches to design, whereas Goel and Pirolli's analysis was concerned with, and based on a cognitively oriented descriptive analysis of design and nondesign activities.

In the domain of cognitive design research, even if often it is not stated explicitly, the tendency is to aim models that generalize across design tasks in different situations—generally equated with task domains. In spite of this more or less implicit adherence to the generic-design hypothesis, and the frequent quotation of Goel and Pirolli's work, the generic-design hypothesis has, except for Goel and Pirolli's work (Goel, 1994; Goel & Pirolli, 1989), received few substantiation through comparative cognitive analyses.

With respect to the difference between design and non-design tasks, one of the exceptions is Lawson's (1979/1984) comparison between students from architecture (design) and from science (non-design),[1] which constitutes one of the bases for the qualification of architects as "solution-focused" and scientists as "problem-focused."

With respect to the similarities between different design tasks, Gross (2003), in his paper *How is a piece of software like a building? Toward general design theory and methods*, advances that pieces of software and buildings are alike on several dimensions. Most will be mentioned below. We have also been focusing on invariants in design activities in different situations. On a rather high level, we have been investigating, for example, if industrial designers did proceed to reuse, as we had observed software designers to do (Visser, 1987)—and we observed indeed interesting forms of reuse by industrial designers participating in the Delft study (Visser, 1995).

---

[1] Lawson's (1979/1984) comparison concerned performances by students (fifth-year architectural and fifth-year science), on artificial tasks supposed to represent architectural-design activities: selecting and arranging coloured blocks of different shapes so as to try to maximize the amount of one certain colour showing around the outside faces, with an undisclosed rule requiring certain blocks to be present.

This paper is a first step in an endeavour to examine the augmented cognitively oriented "generic-design hypothesis" that concludes our book *The Cognitive Artifacts of Designing* (Visser, 2006a). We adhere to the generic-design hypothesis. Yet, we "augment" it, because we suppose that the design situation may introduce specificities in the corresponding design activities, leading to different "forms" of design. Our hypothesis takes the following form:

(1)     Design thinking has distinctive characteristics from other cognitive activities;

(2)     There are commonalities between the implementations of design thinking in different design situations;

(3)     There are also differences between these implementations of design thinking in different situations;

(4)     However, these differences do not reinstate commonalities between designing and other cognitive activities, whereas the commonalities between all the different forms of design thinking are sufficiently distinctive from the characteristics of other cognitive activities, to consider design a specific cognitive activity.

In this paper, we will only be concerned with the third point, that is the different forms of design, and from a descriptive—not normative—cognitive viewpoint. The fourth point remains hypothetical and requires new empirical research comparable to Goel and Pirolli's—but preferentially, in our opinion, performed in professional, that is "real" design situations.

**2. Different Forms of Design**

The idea that there may be different forms of design has been hinted at in informal discussions, generally without empirical or theoretical evidence (Löwgren, 1995; Ullman, Dietterich, & Staufer, 1988). Without any presentation of further underpinning, for example, the engineering-design methodologists Hubka and Eder (1987) assert that "the object of a design activity, what is being designed . . . substantially influences the design process."

It has nevertheless received some empirical confirmatory support. Purcell and Gero (1996), for example, showed a difference between mechanical and industrial designers as regards their susceptibility to "design fixation," that is, people's propensity to take example on possible sources for the artefact to be designed. In "Variants in design cognition," Akin (2001) analyses various design studies and concludes that architects differ from designers in other domains. He resumes the distinctive characteristics in four points: Architects use rich representations, creative design strategies, non-standard problem composition schemata, and complexity management approaches. As a specific feature of architectural design compared to engineering, Akin (2001) advances architects' continuing their search for alternative solutions even if they have already formulated a satisfactory concept: They would not proceed to "premature commitment". In our discussion of design fixation and premature commitment to an early selected kernel idea, we have claimed, however, that the results from cognitive design research on which these affirmations are based, require more analysis (Visser, 2006a).

In addition to the comparison between architecture and other domains, researchers studying software design or HCI have also compared their domain of research to other domains. In the cognitive design research literature, one frequently encounters allusions to, or implicit testimonies of the specific character of software design compared to other types of design—design of HCI is much less the object of discussion in this context. The responsible dimension(s) remain(s), however, unexplored (see Visser, 2006a; 2006b for a discussion of potential specificities).

Some studies have identified examples of variations between design situations other than due to the task domain. In the domain of software design, several authors have observed differences depending on the paradigm and the methodologies involved (e.g., object-oriented vs. procedural, see Lee & Pennington, 1994). We have identified differences between professional designers working on industrial projects and design students or other design knowledgeable participants solving design problems in laboratory experiments (Visser, 1995, 2006a, 2006b).

Advancing our analysis initiated in Visser (2006a; 2006b), this paper proposes six dimensions that we suppose underlie differences between forms of design. We introduce material that still requires further analysis, and indicate a number of directions, for other researchers to follow, modify, complete, and develop. Even if all dimensions are related to the designer, some dimensions mainly depend on them (interindividual differences), whereas others are mainly dependent on the artefact and other situational elements (the organisation of the design process, available tools, design of structures versus design of processes, and artefacts' evolution). As regards the user in the design process, this dimension is rather domain-related, even if certain designers or design teams may of course decide to refer to them in a way that is unusual in their domain (cf. interindividual differences).

2.1 INTERINDIVIDUAL DIFFERENCES

Differences between designers will often affect their representations and activities. This influence may occur by way of one or more of the dimensions proposed hereafter. The use of certain types of representations or other tools, for example, may influence design thinking, but a particular designer may be more inclined to adopt them, or feel more at ease with their use.

*Design Expertise*. A classical cognitive-psychology result confirmed in cognitive design research is that experts and novices in a domain differ as to their representations and activities (Cross, 2004a, 2004b; *Expertise in Design*, 2004).

We have proposed to distinguish, in addition to *levels* of expertise, also different *types* of expertise (Falzon & Visser, 1989; see also Visser & Morais, 1991). We observed how experts in the same domain exhibit (1) different types of knowledge and (2) different organisations of their knowledge—a result comparable to that regarding levels of expertise.

*Routine Character of a Task*. The routine character of people's task is not an objective task characteristic, but depends on the representation people construct of their task. This task characteristic is thus dependent on interindividual differences. It influences, at least in part, the possibilities of reuse in a design project (Visser, 1996). Most design projects comprise both routine and nonroutine tasks. In a comparative analysis of three of our empirical design studies, we have established a link between the more or less routine character of a design project and the way in which analogies are used (at the action-execution and at the action-management levels) (Visser, 1996).

*Idiosyncrasies*. Experts have also been studied in clinical studies, leading researchers to identify idiosyncratic characteristics of particular experts (Cross, 2001, 2002). It is known of Frank Lloyd Wright, e.g., that contrary to most architects, he could conceive of and develop a design entirely without using external representations, not sketching or drawing until the very end of the design process (Weisberg, 1993 quoted in Bilda & Gero, 2005).

"*Personalities*." Among architects, one may encounter the idea that architects differ from other designers because of their "personality." Akin (2001) considers that architects differ from other designers as regards the value they attribute to creative and unique designs.

2.2 THE ORGANISATION OF THE DESIGN PROCESS

The organisation of their task or the process they are involved in, are liable to influence people's activity. Be the organisation imposed by one's hierarchy, or devised by oneself, it works as other tools: It not only structures, but also guides people's activity, through immaterial and material means, such as plans and other methods, representational tools, calculation and simulation aids (cf. the sub-section *Available tools*).

*The Time Scale of the Design Process*. Design is an off-line activity. One might thus naively suppose that, contrary to controllers of dynamic situations, designers have all their time to think over their projects, to analyse and change views, to discuss and confront their views with colleagues. In reality, design generally takes place under temporal constraints—even if their stringency may differ depending on organisational, external (workshop, client), artefactual, and other factors.

*Individual Versus Collective Design*. Certain artefacts are generally designed by an individual designer, others are usually the work of a team. Complexity and size of artefact may play a role, but are certainly not the only variables (they are two dimensions mentioned by Gross, 2003, on which design of software "is like" that of buildings).

We do not see any evidence to suppose that cooperation modifies the nature of the basic cognitive activities and operations implemented in design (i.e., generation, transformation, and evaluation of representations) (Visser, 1993a). Because cooperation proceeds through interaction, it both introduces, however, specific activities and influences designers' representational structures (both on sociocognitive and emotional levels). Some examples of specific activities are coordination, operative synchronization, construction of interdesigner compatible representations, conflict resolution, and management of representations that differ between design partners through confrontation, articulation, and integration. Activities involving argumentation—that is, in our view, activities aiming to modify the representations held by one's interlocutors—obviously play a particularly important role. The construction of interdesigner compatible representations (Visser, 2006a, 2006b), their existence beside designers' private representations, and their management introduce factors that may add complexity to collective design situations compared to individual design.

2.3 THE USER IN THE DESIGN PROCESS

Designers design for other people, the "users" of the artefact. In each domain of design, users are central—even if not always for the designers and even if the use of artefacts may be more or less direct (cf. also *Impact of an Artefact on People's Activity)*. Domains differ, however, with respect to their common practices, for example, concerning the way in which designers usually take into account the potential, future users and their use of the artefact. In design of HCI, for example, there is a tradition and, correspondingly, much effort towards the integration of user data into the design. This has varied from such data being introduced into the design by design participants who "know" the users, but are not these users themselves, to approaches such as participatory design in which the users have themselves a voice in the design process (Carroll, 2006).

It seems likely that the number and variety of participants who take part in a design process influence this process, probably more its socio-organizational than its cognitive aspects. Yet, on a cognitive level, the difficulty of integration may augment with the number of different representations to be integrated—thus with the number of types of participants. In addition, the participation of "nontechnical" design participants may introduce a specific difficulty, both for the users and for their professional design "colleagues."

Gross (2003) mentions two specific user-related dimensions: The difference or equivalence between client and user, and the type of use or user, which may change (more or less), or may remain constant. These two dimensions get a particular weight in the context of the abovementioned influence that number and variety of participants may have on the design process. Notice that for Gross (2003) these are dimensions on which a piece of software is like a building.

2.4 ARTEFACTS: SPATIAL VERSUS TEMPORAL ARTEFACTS

Data concerning the influence of this dimension that concerns "type" of artefact come from results obtained in studies concerning "designing in space versus designing in time".

*Designing in Space Versus Designing in Time*. Studies comparing problems governed by temporal and problems governed by spatial constraints have shown that designers deal differently with these constraints (Chalmé, Visser, & Denis, 2004; see also Thomas & Carroll, 1979/1984). An example of design that preferentially implements temporal constraints is planning (meal planning, see Byrne, 1977; route planning, see Chalmé et al., 2004; Hayes-Roth & Hayes-Roth, 1979). Research, however, has not yet settled clearly the specific relative ease and difficulty involved in the corresponding types of design—it has even less identified the underlying factors.

One might be inclined to generalise this dimension to the opposition structures versus processes. However, structures (which may correspond to states) are not necessarily spatially constrained, even if processes have systematically temporal characteristics. By analogy to the differences between the cognitive treatment of spatial and temporal constraints, one might expect that structures and processes are represented differently (especially mentally, but also externally), thus processed differently, and therefore lead to different design activities (cf. Clancey, 1985's distinction between configuration and planning).

## 2.5 ARTEFACTS' EVOLUTION

"Interactive systems are designed to have a certain behavior over time, whereas houses typically are not," according to Löwgren (1995, p. 94). Even if this assertion is questionable with respect to "behaviour" in general, behaviour over time is a dimension on which artefacts differ (compare, e.g., buildings and interactive software)—and the types of behaviour of different artefacts are quite diverse. An artefact's behaviour over time may be related to its impact on people (the "transformative" nature of artefacts, see Carroll, Rosson, Chin, & Koenemann, 1998), through the interaction that people engage in, and to its use by people who are not necessarily transformed by this use. It may also be due to its deterioration, dependent or independently of people. Two dimensions introduced by Gross (2003) are the degree to which components of the artefact may be subject to change or renewal, and the artefact's lifetime, which may be more or less extended.

All artefacts change over time. Designers are supposed to anticipate the transformation that their artefact products undergo—be it of deterioration or another evolution type. The possibility of anticipation may vary between situations (domains), not necessarily depending on the degree of impact. It depends, among others things, on the possibility of simulating the artefact, or testing it in other ways. For social and interactive artefacts, anticipation may be performed through simulation. The future behaviour of certain technical artefacts may be anticipated based on calculations.

*The Artefact's Impact on People's Activity and the Possibility to Anticipate it*. Predicting people's future use of an artefact product and further anticipating the impact of the artefact on human activity, is one of the "characteristic and difficult properties" of designing (Carroll, 2000, p. 39). Indeed, "design has broad impacts on people. Design problems lead to transformations in the world that alter possibilities for human activity and experience, often in ways that transcend the boundaries of the original design reasoning" (Carroll, 2000, p. 21). Gross (2003) mentions sanitary risks and safety concerns that particular uses or states of an artefact may introduce.

Even if all design has impact on people, certain domains seem more sensitive than others do. HCI, that is the domain with which Carroll (2000, p. 39) is especially concerned in his discussion quoted above, is an example of a domain in which design has particularly broad impacts on people. Yet, this holds for all design with social implications.

*Distance Between Intermediary Representations of the Artefact and of the Product*. The design of an artefact is a different activity than its implementation, but the final representation of an artefact is supposed to be sufficient for its implementation into the artefact product (Visser, 2006a). For certain types of artefacts, however, there seems to be a relatively fluid, steady transition from the different representations of the design concept and the final artefact product—what may be qualified as a shorter "distance" between the two. Software and other symbolic artefacts are an example. This does not imply, however, that, for such artefacts, design and implementation are not distinct. It might, however, clarify our observation that software designers find it particularly difficult to separate design from coding (Visser, 1987).

Based on the distance between concept and product, Löwgren (1995, p. 94) opposes architectural and engineering design to "external" software design ("design of the external behavior and appearance of the product, the services it offers to users and its place in the organization").

*Delay of Implementation*. Design is by definition concerned with artefacts that do not yet exist in the material form of the product. A central aspect of designing is thus, once again, anticipation. The bases of this anticipation may vary depending on other variables (designer's ideas, experience; users' taking part; simulation), but anyhow the conditions of existence, the behaviour, and the use of the artefact products will be more or less different from those anticipated: The world changes without possibility of being completely controlled.

The implementation of certain types of artefacts is much longer in coming than that of others—and not because of laziness or indifference of the designer, the workshop, or the client, or due to lack of resources. Voss, Greene, Post, and Penner (1983) have noticed that the solving of social-science problems is particularly difficult because of the "delay from the time a solution is proposed and accepted to when it is fully implemented" (p. 169). Such a delay clearly complicates the anticipation of the artefact's evolution and other matters involved in its evaluation (though simulation or other means). Even if this observation is particularly applicable to social-science problems, it may also hold for other types of design.

2.6 AVAILABLE TOOLS

In recent years, we have developed the idea that designing is most appropriately qualified as the construction of representations (Visser, 2006a, 2006b). Given this view of design, we privilege representational tools in this discussion, especially external representations and the means to produce them. Designers' internal (mental) representations evidently also play a crucial role in their activity, but these representations are, besides of individual factors, mainly dependent of other components of the situation.

According to Akin (2001), architects differ from designers in other domains with respect to their relative more frequent use of (1) analogue compared to symbolic representations and (2) varying representations. The author attributes this greater variety of representations to architectural design's situated and user-dependent character. Akin also points out the lack of universally accepted representational standards in architecture.

*External Representations.* According to Zhang and Norman (1994), external and internal representations differentially activate perceptual and cognitive processes. With Scaife and Rogers (1996), we presume that things are less systematic, and more complex. Nevertheless, we suppose that the use of internal and that of external representations involve processing differences. Therefore, designing may differ between situations depending on the importance of (certain types of) external representations. One may suppose that, for example, design of physical artefacts (e.g., architectural or mechanical design) differs from design of symbolic artefacts (e.g., procedures or organizations).

1616

Indeed, one of the factors underlying the differences that are often emphasised as existing between software and other types of design may be due to the different types of external representations primarily used. The possibilities provided by particular types of external representations such as sketches and other figurative drawings compared to those offered by alphanumeric representations (especially, with respect to the ease of visualization and manipulation, and their corollaries) may facilitate, for example, simulation and other forms of evaluation of what are going to become physical artefacts.

This observation surely does not only apply to classical (i.e., nonvisual) forms of software design. It probably also holds for other symbolic artefacts, such as other procedures, plans, and organizational structures.

With respect to this dimension, we wish to state explicitly that its importance undeniably also depends on the designer.

*Possible Means for Evaluation*. Domains differ in the methods and other tools that may be used in order to evaluate design proposals (Malhotra, Thomas, Carroll, & Miller, 1980, pp. 129-130). In engineering, more or less objective measures and other criteria for future artefacts' performance can be used and different proposals ranked rather objectively. One can calculate whether a particular design (e.g., a bridge) meets particular functional requirements, such as accommodation and maximum load. The results of qualitative evaluation based on subjective criteria such as aesthetics that are used for other types of design may be more difficult to translate into a score, and thus to compare. In between the extremes of completely objective and entirely subjective evaluation are different types of simulation, physical and mental.

*The Proportion of Reusable Components in the Artefact's Structure* is one of the factors proposed by Gross (2003) that make "a piece of software like a building," but it may differentiate the design of other types of artefacts.

*The Maturity of a Domain* may influence the availability of tools. In 2004, the NSF launched a "Science of Design" program aiming to "develop a set of scientific principles to guide the design of software-intensive systems." An underlying idea was that "in fields more mature than computer science [such as architecture and other engineering disciplines, for example, civil or chemical engineering], design methodology has traditionally relied heavily on constructs such as languages and notational conventions, modularity principles, composition rules, methodical decision procedures and handbooks of codified experience . . . . However, the design of software-intensive systems is more often done using rough guidelines, intuition and experiential knowledge" (Science of Design, 2004).

As noticed above, research in the domain of software design has shown that design methodologies may have an influence on the design activity and on the resulting artefact (Lee & Pennington, 1994). One may suppose that being familiar with the constructs and other tools that have been developed in a domain, may influence, probably facilitate, designers' activity—even if cognitive design research has shown the difficulty of designers' *effectively* working according to design methodology prescriptions (Carroll & Rosson, 1985; Visser & Hoc, 1990).

One may notice that related to the idea that underlies the present dimension and that is only touched upon here, is the question of well-defined versus ill-defined problems and the implications for the nature of the activities involving these problems (see Visser, 2006a, 2006b). From a cognitive-activity viewpoint, most or all ill-defined problems might be analysed as design problems (Visser, 1993b). Going one step further, Falzon (2004) proposes to adopt design as a paradigm for analysing all problem-solving activities. Eventually, Falzon posits, each design problem becomes a state-transformation problem (the typical type of problem examined in classical cognitive psychology), because of people's acquisition of expertise and habits, and of technological evolution. Falzon nevertheless also notes the possibility that there will always remain multiple possible perspectives and situations in which people refuse themselves to refer to procedures and routines. As an example, he refers to a study by Lebahar concerning painters who try to establish conditions that rule out the possibility to refer to routines.

## 3. Conclusion

It is conceivable that not all dimensions mentioned have the same degree of influence on the design activity. Given our view of design as the construction of representations, we might suppose that dimensions related to representational structures and activities are particularly influential. The dimensions advanced are not necessarily independent. They may also depend on other underlying factors and their influence on the activity may exert itself by way of representational structures and activities.

We started this paper noting that we adopted a descriptive cognitively oriented perspective, not a normative one. The dimensions and the characteristics of the different forms of

4activities and cognitive structures may have their methodological and other implications for design support. Given the centrality of representation in designing, the development of appropriate support modalities for representational activities and structures imposes itself. However, according to the role of representation, and the type of representation preferentially used in design tasks in specific design situations, the development of specific support modalities may be worthwhile. Research on these questions may take advantage of the progress already obtained in the domains of software and HCI design. There has been considerable research on visualization and other visual tools, for example, on diagrammatic reasoning (see the diagrammatic reasoning site, retrieved July 06, 2006, from http://zeus.cs.hartford.edu/~anderson/; see also Blackwell, 1997). There is also potentially useful research into representational formats and their exploitation that is not specific to design (e.g., research on multiple —external— representations, see Van Someren, Reimann, Boshuizen, & De Jong, 1988).

Our list of candidate dimensions that might differentiate "forms" of design is a start for further elaboration and analysis. A next step would be to elucidate if indeed, and, if so, how they influence design activity and its result, the artefact. Another direction concerns examination of the fourth point of our hypothesis, namely that in spite of the possibly different forms of design, design remains a specific cognitive activity relative to other cognitive activities.

**References**

Akin, Ö. (2001). Variants in design cognition. In C. Eastman, M. McCracken & W. Newstetter (Eds.), *Design knowing and learning: Cognition in design education* (pp. 105-124). Amsterdam: Elsevier Science.

Bilda, Z., & Gero, J. S. (2005). Do we need CAD during conceptual design? In B. Martens & A. Brown (Eds.), *Computer Aided Architectural Design Futures 2005. Proceedings of the 11th International CAAD Futures Conference held at the Vienna University of Technology, Vienna, Austria, on June 20-22, 2005* (pp. 155-164). London: Springer.

Blackwell, A. F. (1997). *Diagrams about thoughts about thoughts about diagrams*. Retrieved June 29, 2006, from http://www.cl.cam.ac.uk/~afb21/publications/AAAI.html

Byrne, R. (1977). Planning meals: Problem-solving on a real data-base. *Cognition, 5*, 287-332.

Carroll, J. M. (2000). *Making use. Scenario-based design of human computer interactions*. Cambridge, MA: MIT Press.

Carroll, J. M. (2006). Dimensions of participation in Simon's design. *Design Issues, 22*(2), 3-18.

Carroll, J. M., & Rosson, M. B. (1985). Usability specifications as a tool in iterative development. In H. R. Hartson (Ed.), *Advances in human-computer interaction* (Vol. 1, pp. 1-28). Norwood, NJ: Ablex.

Carroll, J. M., Rosson, M. B., Chin, G., & Koenemann, J. (1998). Requirements development in scenario-based design. *IEEE Transactions on Software Engineering*, 24(12), 1156–1170.

Chalmé, S., Visser, W., & Denis, M. (2004). Cognitive effects of environmental knowledge on urban route planning strategies. In T. Rothengatter & R. D. Huguenin (Eds.), *Traffic and transport psychology. Theory and application* (pp. 61-71). Amsterdam: Elsevier.

Chandrasekaran, B. (1983). Towards a taxonomy of problem solving types. *AI Magazine, 4 (Spring)*(1), 9-17.

Clancey, W. J. (1985). Heuristic classification. *Artificial Intelligence, 27*, 289-350.

Cross, N. (2001). Strategic knowledge exercised by outstanding designers. In J. S. Gero & K. Hori (Eds.), *Strategic knowledge and concept formation III* (pp. 17-30). Sydney, Australia: University of Sydney, Key Centre of Design Computing and Cognition.

Cross, N. (2002). Creative cognition in design: Processes of exceptional designers. In T. Kavanagh & T. Hewett (Eds.), *Creativity and Cognition 2002 (C&C'02, the 4th conference on Creativity & Cognition)* (pp. 14-19). New York: ACM Press.

Cross, N. (2004a). Expertise in design. Introduction. *The Journal of Design Research, 4*(2).

Cross, N. (Ed.). (2004b). Expertise in design [Special issue]. *Design Studies, 25*(5).

Expertise in Design [Special issue]. (2004). *The Journal of Design Research, 4*(2).

Falzon, P. (2004). Préface [Preface]. In P. Falzon (Ed.), *Ergonomie* [Ergonomics] (pp. 11-13). Paris: Presses Universitaires de France.

Falzon, P., & Visser, W. (1989). Variations in expertise: Implications for the design of assistance systems. In G. Salvendy & M. J. Smith (Eds.), *Designing and using human-computer interfaces and knowledge based systems* (Vol. II, pp. 121-128). Amsterdam: Elsevier.

Goel, V. (1994). A comparison of design and nondesign problem spaces. *Artificial Intelligence in Engineering, 9*, 53-72.

Goel, V., & Pirolli, P. (1989). Motivating the notion of generic design within information-processing theory: The design problem space. *AI Magazine, 10*(1), 18-36.

Gross, M. D. (2003, November 2-4). *How is a piece of software like a building? Toward general design theory and methods*. Paper presented at the National Science Foundation (NSF) Invitational Workshop on Science of Design: Software Intensive Systems, Airlie Center, Virginia.

Hayes-Roth, B., & Hayes-Roth, F. (1979). A cognitive model of planning. *Cognitive Science, 3*, 275-310.

Hubka, V., & Eder, W. E. (1987). A scientific approach to engineering design. *Design Studies, 8*(3), 123-137.

Lawson, B. R. (1979/1984). Cognitive strategies in architectural design. In N. Cross (Ed.), *Developments in design methodology* (pp. 209-220). Chichester, England: Wiley (Originally published in *Ergonomics*, 1979, *22* (1), 59-68.).

Lee, A., & Pennington, N. (1994). The effects of paradigms on cognitive activities in design. *International Journal of Human-Computer Studies, 40*, 577-601.

Löwgren, J. (1995). Applying design methodology to software development. In G. M. Olson & S. Schuon (Eds.), *Symposium on Designing Interactive Systems. Proceedings of the conference on Designing interactive systems: processes, practices, methods, and techniques (DIS'95)* (pp. 87-95). New York: ACM Press.

Malhotra, A., Thomas, J. C., Carroll, J. M., & Miller, L. A. (1980). Cognitive processes in design. *International Journal of Man-Machine Studies, 12*, 119-140.

Newell, A., & Simon, H. A. (1972). *Human problem solving*. Englewood Cliffs, NJ: Prentice-Hall.

Purcell, T., & Gero, J. S. (1996). Design and other types of fixation. *Design Studies, 17*(4), 363–383.

Scaife, M., & Rogers, Y. (1996). External cognition: How do graphical representations work? *International Journal of Human-Computer Studies*, 45(1), 185-213.

Science of Design. (2004, September 6, 2005, last update date). Retrieved October 12, 2005, from http://www.nsf.gov/pubs/2004/nsf04552/nsf04552.htm

Thomas, J. C., & Carroll, J. M. (1979/1984). The psychological study of design. *Design Studies*, 1(1), 5-11. Also in Cross, N. (Ed.). (1984). *Developments in design methodology* (pp. 1221-1235). Chichester, United Kingdom: Wiley.

Ullman, D. G., Dietterich, T. G., & Staufer, L. A. (1988). A model of the mechanical design process based on empirical data. *AI EDAM, 2*, 33-52.

Van Someren, M. W., Reimann, P., Boshuizen, H. P. A., & De Jong, T. (1988). *Learning with multiple representations*. Amsterdam: Elsevier.

Visser, W. (1987). Strategies in programming programmable controllers: A field study on a professional programmer. In G. M. Olson, S. Sheppard & E. Soloway (Eds.), *Empirical Studies of Programmers: Second workshop (ESP2)* (pp. 217-230). Norwood, NJ: Ablex.

Visser, W. (1993a). Collective design: A cognitive analysis of cooperation in practice. In N. F. M. Roozenburg (Ed.), *Proceedings of ICED 93, 9th International Conference on Engineering Design* (Vol. 1, pp. 385-392). Zürich, Switzerland: HEURISTA.

Visser, W. (1993b). Design & knowledge modeling - Knowledge modeling as design. *Communication and Cognition - Artificial Intelligence, 10*(3), 219-233.

Visser, W. (1995). Use of episodic knowledge and information in design problem solving. *Design Studies, 16*(2), 171-187.
also in N. Cross, H. Christiaans & K. Dorst (Eds.) (1996), *Analysing design activity* (Ch. 1913, pp. 1271-1289). Chichester, United Kingdom: Wiley.

Visser, W. (1996). Two functions of analogical reasoning in design: A cognitive-psychology approach. *Design Studies, 17*, 417-434.

Visser, W. (2006a). *The cognitive artifacts of designing*. Mahwah, NJ: Lawrence Erlbaum Associates.

Visser, W. (2006b). Designing as construction of representations: A dynamic viewpoint in cognitive design research. *Human-Computer Interaction, Special Issue "Foundations of Design in HCI", 21*(1), 103-152.

Visser, W., & Hoc, J.-M. (1990). Expert software design strategies. In J.-M. Hoc, T. Green, R. Samurçay & D. Gilmore (Eds.), *Psychology of programming* (pp. 235-250). London: Academic Press.

Visser, W., & Morais, A. (1991). Concurrent use of different expertise elicitation methods applied to the study of the programming activity. In M. J. Tauber & D. Ackermann (Eds.), *Mental models and human-computer interaction* (Vol. 2, pp. 59-79). Amsterdam: Elsevier.

Voss, J. F., Greene, T. R., Post, T. A., & Penner, B. C. (1983). Problem-solving skill in the social sciences. In G. Bower (Ed.), *The psychology of learning and motivation* (Vol. 17, pp. 165-213). New York: Academic Press.

Zhang, J., & Norman, D. A. (1994). Representations in distributed cognitive tasks. *Cognitive Science, 18*, 87-122.